\documentclass[12pt,preprint]{aastex}
\shorttitle{Spitzer Space Telescope Mission}
\shortauthors{Werner et al.}

\slugcomment{To be submitted to the Astrophysical Journal Supplement - Spitzer Issue}

\begin{document}
\title{The Spitzer Space Telescope Mission}

\author{M.~W.~Werner,\altaffilmark{1}
T.~L.~Roellig,\altaffilmark{2}
F.~J.~Low,\altaffilmark{3} G.~H.~Rieke,\altaffilmark{3} M.~Rieke,\altaffilmark{3}
W.~F.~Hoffmann,\altaffilmark{3} E.~Young,\altaffilmark{3}
J.~R.~Houck,\altaffilmark{4} B.~Brandl,\altaffilmark{5}
G.~G.~Fazio,\altaffilmark{6} J.~L.~Hora,\altaffilmark{6}
R.~D.~Gehrz,\altaffilmark{7}
G.~Helou,\altaffilmark{8} B.~T.~Soifer,\altaffilmark{8} 
J.~Stauffer,\altaffilmark{8} J.~Keene,\altaffilmark{8,9}
P.~Eisenhardt,\altaffilmark{9} D.~Gallagher,\altaffilmark{9}
T.~N.~Gautier,\altaffilmark{9} W.~Irace,\altaffilmark{9} C.~R.~Lawrence,
\altaffilmark{9} L.~Simmons,\altaffilmark{9}
J.~E.~Van~Cleve,\altaffilmark{10}
M.~Jura,\altaffilmark{11} E.~L.~Wright,\altaffilmark{11} and
D.~P.~Cruikshank\altaffilmark{2}}

\altaffiltext{1}{Jet Propulsion Laboratory, Mailstop 264-767, 4800 Oak Grove 
Dr., Pasadena, CA 91109, mwerner@sirtfweb.jpl.nasa.gov}
\altaffiltext{2}{NASA Ames Research Center, MS 245-6, Moffett
  Field, CA 94035-1000}
\altaffiltext{3}{Steward Observatory, University of Arizona, 933 N. 
Cherry Ave., Tucson, AZ 85721}
\altaffiltext{4}{Cornell University, Astronomy Department, Ithaca, NY 14853-6801}
\altaffiltext{5}{Leiden University, 2300 RA Leiden, The Netherlands}
\altaffiltext{6}{Harvard-Smithsonian Center for Astrophysics, 
60 Garden St., 
Cambridge, MA 02138}
\altaffiltext{7}{Department of Astronomy,
School of Physics and Astronomy,
116 Church Street, S. E.,
University of Minnesota,
Minneapolis, Minnesota 55455}
\altaffiltext{8}{Spitzer Science Center, California Institute of Technology, 
Pasadena, CA 91125}
\altaffiltext{9}{Jet Propulsion Laboratory, 4800 Oak Grove Dr., Pasadena, CA 91109}
\altaffiltext{10}{Ball Aerospace and Technologies Corp., 1600 Commerce 
St., Boulder, CO 80301}
 \altaffiltext{11}{Department of Physics and Astronomy, University
of California, Los Angeles, CA 90095-1562}

\begin{abstract}

The Spitzer Space Telescope, NASA's Great Observatory for infrared astronomy, 
 was launched 2003 August 25 and is returning excellent scientific data from its 
 Earth-trailing solar orbit.  Spitzer combines the intrinsic sensitivity achievable 
 with a cryogenic telescope in space with the great imaging and spectroscopic 
 power of modern detector arrays to provide the user community with huge gains 
 in capability for exploration of the cosmos in the infrared.  The observatory systems are 
 largely performing as 
 expected and the projected cryogenic lifetime is in excess of 5 years. This paper summarizes 
 the on-orbit scientific, technical and operational performance of Spitzer.  
 Subsequent papers in this special issue describe the Spitzer instruments in 
 detail and highlight many of the exciting scientific results obtained during the 
 first six months of the Spitzer mission.

\end{abstract}


\section{Introduction}

The Spitzer Space Telescope, NASA's Great Observatory for infrared astronomy, 
 was launched 2003 August 25 and is performing extremely well and returning 
 excellent scientific data from its Earth-trailing solar orbit.  Spitzer incorporates 
 an 85 cm diameter telescope primary mirror, cooled to as low as 5.5 K, and 
 three scientific instruments providing imaging and spectroscopy at wavelengths 
 from 3.6 to 160 $\mu$m. The observatory systems are largely performing as 
 expected and the projected cryogenic lifetime is in excess of 5 years. The 
 ground systems and operations teams at JPL, Lockheed Martin Astronautics, 
 and the Spitzer Science Center are routinely downloading 1.1 GBytes of data 
 per day and delivering the pipeline-processed science data to the observer 
 within 14 days.  This paper provides an overview of the observatory design 
 and performance, emphasizing the pointing, optical, and cryo-thermal 
 systems which are of immediate concern to the Spitzer user community.  
 We also present summaries of the instrumental capabilities and sensitivities, 
 and of the orbit and operations.  The results presented in companion papers 
 in this special issue show that Spitzer will provide both greatly increased 
 understanding of known astrophysical questions and discoveries which 
 define new areas of scientific inquiry.  At the same time, Spitzer demonstrates 
 a number of observatory and mission design innovations which can be 
 productively incorporated into future space programs. Current details on the 
 Spitzer status and performance can be found on the Spitzer Science Center 
 web site.\footnote{http://ssc.spitzer.caltech.edu}

\section{Satellite Design}

The Spitzer flight hardware consists of a spacecraft that operates at roughly 
  ambient temperature and a Cryogenic Telescope Assembly (CTA) that is 
 cooled by a combination of superfluid liquid helium, helium boil-off gas, 
 and radiative cooling and operates at much lower temperatures. The spacecraft 
 was built by Lockheed Martin, Sunnyvale, and the CTA by Ball Aerospace.  
 The two components are thermally isolated from each other with a system of 
 low-thermal conductivity struts and shields. The overall layout of the flight 
 configuration is shown in Figure~1, while summary information about the  
 complete flight system is given in Table~1. More detailed descriptions of the 
 two components are given below.

\noindent{\it The Spacecraft---}The Spitzer spacecraft handles the 
observatory's power
generation, pointing and momentum control, data processing and storage, and
telecommunications functions. It also contains the warm electronics 
portions of the three
scientific instruments. The spacecraft structure itself is octagonally shaped, 
constructed from graphite
composite and aluminum honeycomb material, and is mounted to the cold CTA with
low-thermal-conductivity gamma-alumina struts. A set of aluminized 
Mylar thermal blankets and
aluminum/graphite epoxy honeycomb radiation shields dramatically 
reduces the radiative thermal
load from the warm spacecraft onto the CTA.  A similar system of thermal 
shields attenuates the
thermal radiation from the solar array, which is cantilevered from 
the spacecraft
structure without any physical connection to the cold CTA.  There are approximately 1,500 wires 
 that run from the warm spacecraft to the cold CTA. These were constructed of very low thermal-conductivity materials and carefully heat-sunk so that they caused minimal parasitic heat input.
 The spacecraft avionics are 
 fully redundant, as are the 
components of the
Pointing and Control System (PCS) and Reaction Control System (RCS) 
discussed below.  

The PCS comprises reaction wheels, gyroscopes, and star trackers. 
Four reaction wheels
mounted in a pyramid arrangement provide redundancy against a single 
wheel failure.  As the spacecraft is well outside
the Earth's magnetic field, accumulated angular momentum is removed through an
ambient-temperature, pressurized-nitrogen gas \hbox{RCS}.  The 
nitrogen gas plume does not
affect the infrared environment and the sensitive CTA thermal control 
surfaces.  Twice a day a redundant
pair of visible light sensors (Pointing Calibration Reference Sensors; PCRS) 
in the telescope focal plane 
provide
calibration of the telescope bore-sight to the star trackers, which 
are mounted to the
spacecraft structure.

\noindent{\it The Cryogenic Telescope Assembly---}The CTA consists of 
the telescope
assembly, the cryogenic portions of the three science instruments, a 
superfluid helium
dewar, and various thermal shields.  Spitzer employs a novel thermal 
design (hereafter referred to as the ``warm launch'' design) in which most 
of the mass of the CTA, including the telescope and
its baffles, is external to the cryostat 
vacuum shell and is launched at ambient
temperature and only begins cooling when on-orbit.  This allows for a much 
smaller vacuum pressure
vessel and a smaller total observatory mass than would a design based on 
the more conventional ``cold launch'' architecture used in the earlier
IRAS and ISO missions.  The components of the CTA external to the 
cryostat are cooled after
launch by a combination of radiation and helium boil-off vapor under 
a concept developed by
F. Low and described in more detail in \citet{lys95}.\footnote{Other possible applications
of radiative cooling to the design of an infrared space observatory are described in
\cite{bel92}, and alternate configurations were also proposed by H. Moseley at about
the same time. }  A very low-temperature CTA
outer shell is critical to
the achievement of a long mission lifetime with a small cryostat, but requires
excellent thermal isolation between warm and cold components as well 
as tight restrictions on
the orientation of the flight hardware with respect to the Sun.  
 The CTA outer shell is painted with high-emissivity black 
paint on the side that
faces away from the Sun and is polished aluminum on the sides that 
face the thermal shields,
solar array, and warm spacecraft components. The Spitzer warm launch design
 and the wide range of infrared wavelengths observed by the focal plane instruments
 require an open aperture in the vacuum cryostat after
 launch to allow the passage of infrared light from the telescope into the instrument
 chamber. For this purpose a  removable vacuum window was constructed of gold film-coated
 sapphire to allow the cold instruments to perform visible and near-infrared alignment of the CTA optical
 train on the ground without having to deploy the window prior to launch. A dust cover 
 over the outer shell aperture that was ejected
4.9 days after launch, and 0.9 days prior to the opening of the cryostat window,
 was the only other deployable mechanism on the Spitzer observatory.

The telescope optics and metering structure are constructed entirely 
of beryllium, minimizing
changes in both the telescope prescription and its alignment with the 
focal plane as the
telescope cools on-orbit.  A moveable secondary capable of motion only along the
optical axis was installed to 
compensate for any changes
in focus due to gravity release and uncertainties in the ground 
testing of the telescope.
The telescope optics are not cooled directly by the helium bath but 
are thermally coupled to
the cryostat vacuum shell, which is cooled by the superfluid helium 
boil-off vapor.  The
higher the boil-off rate, the colder the telescope.  The telescope 
temperature necessary to
keep thermal emission from the optics negligible compared to the 
natural zodiacal background
decreases with wavelength, reaching 5.6\,K for the  160\,$\mu$m channel in the 
 Multiband Infrared Photometer for Spitzer (MIPS) instrument.  MIPS 
itself does not dissipate
enough power in its cold assemblies to provide the necessary boil-off
rate for this temperature; a small adjustable heater in the cryogen 
 provides a few extra milliwatts of power if needed.  The
helium liquid-gas interface is constrained within a porous plug phase separator,
preventing rapid helium
coolant loss.

\section{Instrument Payload}

Three infrared instruments, the Infrared Array Camera (IRAC), the Infrared
Spectrograph
 (IRS), and the MIPS, 
share a common focal
plane.  Their fields of view are defined by pickoff mirrors, as shown 
in Figure~2.  The top
level design and performance characteristics of the instruments are 
described in Tables~2--4.
The instruments achieve great scientific power through
the use of state-of-the-art infrared detector arrays in formats as 
large as $256\times256$
pixels.  For broadband imaging and low spectral resolution 
spectroscopy, Spitzer has achieved
sensitivities close to or at the levels established by the natural 
astrophysical
backgrounds---principally the zodiacal light---encountered in Earth 
orbit (Figure 3).  The only moving
part in use in the entire science instrument payload is a scan mirror 
in the MIPS. The instrument aperture is selected by body pointing the 
 entire observatory.

\section{Orbit}

Spitzer utilizes a  unique Earth-trailing solar orbit (as of 26 March 
2004 Spitzer was
trailing 0.083 AU  behind the Earth).  As seen from Earth, Spitzer 
recedes at about
0.12\,AU\,yr$^{-1}$ and will reach a distance of 0.62\,AU in five 
years.  The Earth-trailing
orbit has several major advantages over a low-Earth orbit for 
Spitzer.  The principal
advantage is being away from the heat of the Earth; this enables the warm launch 
architecture
and the extensive use of radiative cooling which makes Spitzer's 
cryo-thermal design
extremely efficient.  More precisely, the solar orbit allows the 
spacecraft always to be
oriented with the solar array pointed at the Sun while the black side of the 
 outershell has a complete hemispherical
view of deep space with no intefering heat sources, enabling the 
 radiative cooling  of the
 outer shell described earlier.  
The absence of eclipses
in the solar orbit makes for an extremely stable thermal 
configuration and reduces
variability of the alignment between the cold telescope and the warm 
star tracker and
inertial reference unit (gyroscopes) to less than 0.5 arc-seconds over 
timescales of days.
Operationally, the orbit permits excellent sky viewing and observing 
efficiency  \citep{kwo93}.  Finally,
while in a solar orbit the observatory is not affected by the charged 
particles in the Van
Allen radiation belts.  These thermal and charged particle 
advantages are shared by the $L_2$ orbit
contemplated for \hbox{JWST}.  Unlike an $L_2$ orbit, however, 
Spitzer's heliocentric orbit
eliminates the need for station-keeping, allowing the use of a smaller and 
less-costly launch vehicle.

The orbit has two disadvantages.  Firstly, as Spitzer moves away from 
the Earth, the power
margins for data transmission and command communication decrease, forcing 
changes in communication
strategy and, eventually, somewhat reduced efficiency as more time is 
spent communicating with the ground stations.
The data are transmitted to the antennae of the Deep Space Network 
by orienting the S/C so
that a fixed X-band antenna located on the bottom of the spacecraft 
is pointed at the
Earth.  Our current data transmission strategy of two $\sim$45-minute passes 
per day at a
data rate of 2.2\,Mbps is robust for at least the first 2.5~years of 
the mission  Secondly,
outside of the protection of the Earth's magnetosphere Spitzer is more 
susceptible to solar proton
storms and is subject to a higher quiescent galactic 
cosmic rate flux.  However,
as mentioned above Spitzer is also free from the daily passages 
through the South
Atlantic Anomaly which provide the main radiation dose in Low Earth Orbit.

During its In-Orbit Checkout period Spitzer 
encountered a very large solar
proton storm beginning 2003 October 28.  During the two-day period of 
this storm the
observatory received an integrated dose of $1.6 \times 
10^9$\,p$^+$\,cm$^{-2}$ for proton
energies greater than 50\,MeV, roughly the integrated dose expected 
under normal conditions
over the first 2.5~years of operations.  During the course of the storm
the science instruments were powered off.
Aside from a small number ($<4$\%) of IRS detector pixels that 
exhibited increased dark
current, a correctible change in the offset bias of the wide-angle sun sensors 
on the S/C, and a very slight decrease in the efficiency of its solar arrays, the
observatory was unaffected by this event.

\section{Operations}

The users' operational interface to the observatory is through the 
Spitzer Science Center
[SSC] at Caltech, which is responsible for science program selection 
and scheduling,
calibration, pipeline processing, and data distribution and archiving.  The SSC
collaborates with mission and spacecraft operations teams at JPL and 
Lockheed Martin,
Denver, in preparing and executing the on-orbit sequences.  

The underlying principle 
of the science operations 
is that of the
Astronomical Observational Template, a web-based menu which the user 
fills in to define a
particular observation.  A filled in AOT becomes an Astronomical Observational Request, or 
\hbox{AOR}.  
The AOR created by the observer is 
expanded directly into a
series of spacecraft commands, and the executed sequence consists of 
a series of AORs, each
individually lasting between 10~minutes and six hours, interspersed 
with spacecraft
activities such as pointing system calibrations and data transmission. 

Only one instrument operates at a 
time; there are no
parallel observations or internal calibrations, greatly 
simplifying the onboard
architecture and software.  Because the 
helium utilization is
dominated by the instrument cold power dissipation, operating two 
instruments at once would
halve the lifetime even as the instantaneous data rate doubled.  

With the large fraction ($\sim$35\%) of the sky that is accessible at any time, Spitzer can 
be operated with high efficiency.  Our
target is to achieve an on-orbit efficiency (defined as $1-f_c$, 
where $f_c$ is the the
fraction of wall-clock time spent in spacecraft calibration and data 
transmission) of 0.9.
Pre-launch simulations suggested that the efficiency could exceed 
0.88 on orbit, and early
results from the first few months of normal operations show that 
we should reach the
0.9 level as our operational experience accumulates.

\section{Observing Time Utilization}

Observing time on Spitzer is divided among three major categories of 
users:  General
Observers (GO), Legacy Science Teams, and Guaranteed Time Observers (GTO). 
The GTOs receive 20\% 
of the time for the first 2.5 years and 15\% thereafter. In 
addition, up to 5\% of
the observing time can be allocated as Director's Discretionary Time while the balance of the
time is available for GO use.   Six Legacy 
Science teams were 
selected prior to
launch to carry out large-scale
projects with the joint objectives of creating a coherent scientific 
legacy and seeding the
Spitzer archive with data that will stimulate follow-on proposals 
from the entire scientific
community.   Details of the presently planned GTO and Legacy programs can 
be found on the SSC web site.
The GO time constitutes
approximately 75\% of the available observing time and is available to the 
worldwide scientific
community through the usual peer-review process and will be 
allocated on an annual basis. GO Cycle 1 proposals were submitted in 2004 February 
and Cycle 1 observations will began in 2004 July.  For the next cycle, with proposals 
due in 2005 February, approximately
3800 hours of GO observing time will be allocated.

\section{On-Orbit Status}

Spitzer requires the unique low thermal environment of space in order to
function properly; the environment both permits the necessary 
radiative cooling and
provides the low backgrounds to allow the science instruments to 
achieve their high
sensitivities. This environment was difficult or impossible to 
simulate on the ground so
that there was some uncertainty in how well the observatory would 
perform in orbit. We are
pleased to report that in almost every respect the on-orbit 
performance of the observatory
meets or exceeds expectations.  The on-orbit performance of the 
instruments was summarized
above and more details can be found in \citet{faz04}, \citet{hou04}, and \citet{rie04}. 
Aside from the science instrument sensitivities, the most 
important on-orbit
performance characteristics are listed below:

\noindent{\it Thermal---}The outermost CTA shield - the outer shell - attained its
final temperature of 34-34.5 K solely by radiative cooling, and the telescope cooled 
to its operating temperature of 5.6 K in 41 days. The mass of the helium remaining after the
initial cooldown was measured to be 43.4\,kg  by applying a heat pulse
to the helium bath 57~days after launch and measuring the accompanying temperature rise. 
To maintain the telescope at 5.5 K requires a total heat of 5.6 mW to the helium bath,
leading to a boil-off of approximately 22 g of helium per day. For a fixed telescope temperature, these
nominal numbers translate into a 5.3 year post-launch lifetime.
In a worst-case scenario where all of the engineering 
uncertainties stack against us,
the lifetime would drop to 4.0~years, which still compares quite 
favorably with the mission
minimum lifetime requirement of 2.7~years.  By  regulating the power dissipation to 
accommodate the telescope
temperature requirements of the instrument that is in use, we are 
expecting at least an
additional four months of cryogenic lifetime beyond the nominal 
5.3~years. The only other expendable, the $\rm N_2$ gas in the RCS system, should last for 
more than 10 years.  Following the
depletion of the cryogens, the telescope will warm up, but it will 
still be colder than the
outer shell.  Current estimates suggest that the telescope temperature will always be 
$<30$\,\hbox{K}.  At this
temperature, both the instrumental background and the detector dark 
current should be low
enough for Spitzer to continue natural background-limited operations 
in the shortest
wavelength IRAC bands at 3.6 and 4.5\,$\mu$m.  Additional information 
on the measured
on-orbit thermal performance of the Spitzer CTA can be found in \citet{fin04}.

\noindent{\it Optical performance, including focus---}A focus campaign of two 
secondary mirror movements was initiated 
once the telescope
assembly had cooled to operating temperature to achieve the final focus \citep{hof04}. 
On-orbit 
measurements of the
telescope then showed that it provides diffraction-limited performance at 
wavelengths greater than
5.5\,$\mu$m, which compares favorably with the requirement of 
6.5\,$\mu$m \citep{geh04}.   
After the completion of
this campaign all of the science instruments were measured to be 
confocal to within the
depths of their focus.

\noindent{\it Pointing performance---}In general, the pointing 
performance of Spitzer is
better than the nominal predictions. The star tracker has proven to 
be very accurate, with a
noise-equivalent-angle of approximately 0.11~arcseconds using an 
average of 35 tracked
stars. The cryo-mechanical 
variation in alignment
over time between the star tracker mounted on the warm spacecraft 
and the cold
telescope bore-sight has proven to be very small and easily 
calibrated, so that the star
tracker can be used to directly point the telescope boresight to 
better than 1 arc-second,
1-$\sigma$ RMS radial uncertainty. For offset movements less than 30~arcminutes, the 
PCS meets the
0.4~arcsecond offset accuracy requirement.  It takes less than 150 seconds for the telescope 
to slew 15 degrees and to settle to its commanded position to within the above tolerances. 
Once the telescope pointing has settled, it is stable to
within 0.03~arc-second, 1-$\sigma$ RMS radial uncertainty, for times up to 600~seconds. Observations of 
a number of solar system targets demonstrated that the
PCS is able to track such objects at rates up to 0\farcs14 per second. The PCS has also demostrated smooth scanning during MIPS scan-mapping observations at rates of 2, 6, and 20 arc-seconds per second.

\noindent{\it External torques and reaction control---}Net angular 
momentum will build up
on orbit from a combination of solar pressure and uncompensated helium 
venting. During the design of
the observatory care was taken to align the vector of solar pressure 
with the center of
mass as closely as possible, and the two CTA helium vent nozzles were also balanced very well. This has
resulted in very low momentum build-up and a low frequency---once or 
twice per week---of
actuations of the reaction control system to remove angular momentum.

\section{Summary and Conclusions}

The Spitzer Space Telescope has been operating at high efficiency since December, 2003.  The early scientific 
results presented in this special volume, based largely on observations made prior to March, 2004, are indicative 
of Spitzer's great power and promise.   We welcome the participation of the entire scientific community over the 
next five years in using this important new facility to expand yet further our understanding of the Universe.

\section{Acknowledgments}

The Spitzer Space Telescope is operated by the Jet Propulsion 
Laboratory, California
Institute of Technology under NASA contract 1407. Support for this 
work was provided by
NASA through an award issued by JPL/Caltech. T. Roellig would like to 
acknowledge the
support of the NASA Office of Space Sciences.

The success of the Spitzer Space Telescope reflects the talent and dedication of thousands 
of people who have worked on this project over the past two+ decades.  The authors of 
this paper consider it a privilege to recognize our colleagues who have contributed to this 
success.  In the following we list many of these people, sorted by the institutions through 
which they participated in the Spitzer definition, development and/or early operations. 
We hope that this recognition provides a tangible though modest token of our gratitude 
and that they Ð and others whom we may have inadvertently omitted from the lists - are 
able to share with us the excitement of the first fruits of our collective labors.

{\bf  Ball Aerospace Ð Cryogenic Telescope Assembly:} {\footnotesize \it R. Abbott, D. Adams, S. Adams, J. Austin, 
B. Bailey, H. Bareiss, J. Barnwell, T. Beck, B. Benedict, M. Bilkey, W. Blalock, M. Breth, R. 
Brown, D. Brunner, D. Burg, W. Burmester, S. Burns, M. Cannon, W. Cash, T. Castetter, M. 
Cawley, W. Cebula, D. Chaney, G. J. Chodil, C. Cliff, S. Conley, A. Cooper, J. Cornwell Sr, L. Cortelyou, J. Craner, K. 
Craven, D. Curtis, F. Davis, J, Davis, C. Dayton, M. Denaro, A. DiFronzo, T. Dilworth, N. 
Dobbins, C. Downey, A. Dreher, R. Drewlow, B. Dubrovin, J. Duncan, D. Durbin, S. Engles, P. 
Finley, J. Fleming, S. Forrest, R. Fredo, K. Gause, M. Gee, S. Ghesquiere, R. Gifford, J. Good, 
M. Hanna, D. Happs, F. Hausle, G. Helling, D. Herhager, B. Heurich, E. Hicks, M. Hindman, R. 
Hopkins, H. Hoshiko Jr, J. Houlton, J. Hueser, J. Hurt, W. Hyatt, K. Jackson, D. Johnson, G. 
Johnson, P. Johnson, T. Kelly, B. Kelsic, S. Kemper, R. Killmon, R. Knewtson, T. Konetski, B. Kramer, R. 
Kramer, L. Krauze, T. Laing, R. LaPointe, J. Lee, D. Lemon, P. Lien, R. Lytle, L. Madayev, M. Mann, R. 
Manning, J. Manriquez, M. Martella, G. Martinez, T. McClure, C. Meier, B. Messervy, K. 
Modafferi, S. Murray, J. Necas Jr, M. Neitenbach, P. Neuroth, S. Nieczkoski, G. Niswender, E. 
Norman-Gravseth, R. Oonk,L. Oystol, J. Pace, K. Parrish, A. Pearl Jr, R. Pederson, S. 
Phanekham, C. Priday, B. Queen, P. Quigley, S. Rearden, M. Reavis, M. Rice, M. Richardson, P. 
Robinson, C. Rowland, K. Russell, W. Schade, R. Schildgen, C. Schroeder, G. Schultz, R. 
Schweickart, J. Schweinsberg, J. Schwenker, S. Scott, W. Seelig, L. Seide, K. Shelley, T. Shelton, 
J. Shykula, J. Sietz, J. Simbai, L. Smeins, K. Sniff, B. Snyder, B. Spath, D. Sterling, N. Stoffer, B. 
Stone, M. Taylor, R. Taylor, D. Tennant, R. Tio, P. True II, A. Urbach, S. Vallejo, K. Van Leuven, 
L. Vernon, S. Volz, V. VonRuden, D. Waldeck, J. Wassmer, B. Welch, A. Wells, J. Wells, T. 
Westegard, C. Williamson, E. Worner Jr,T. Yarnell, J. Yochum, A. Youmans, J. Zynsky.}

{\bf Cornell University - Infrared Spectrograph:} {\footnotesize \bf Cornell (Project Mgmt. \& Science)} { \footnotesize
\it D. Barry, S. V. W. Beckwith, J. 
Bernhard-Salas, C. Blacken,  V. Charmandaris, M. Colonno, S. Corbin, P. Devine, D. Devost, J. 
Diller, K. Duclos, E. Furlan, G. Gull, P. Hall, L. Hao, C. Henderson, T. Herter, J. Higdon, S. Higdon, 
 P. Howell, L. McCall, A. Parks, B. Pirger, A. Rakowski, S. Reinehart, A. Reza, E. E. Salpeter, 
 J. Schoenwald, G. 
Sloan, J. Smith, H. Spoon, K. Uchida,  D. Weedman, J. Wilson} 
{\footnotesize \bf University of Rochester} { \footnotesize \it D. M. Watson, 
W.F. Forrest} {\footnotesize \bf California Institute of Technology}  
{ \footnotesize \it K. Matthews}
{\footnotesize \bf Ball Aerospace (Instrument 
Development)} { \footnotesize \it D. Alderman, D. Anthony, M. Bangert, J. Barnwell, A. Bartels, S. Becker, W. Belcher, J. 
Bergstrom, D. Bickel, M. Bolton,
S. Burcar, D. Burg, S. Burns, S. Burns, D. Burr, P. Burrowes, W. Cebula, C. Conger, J. Crispin, M. Dean, 
M. D'Ordine, S. Downey, R. Drewlow, L. Duchow, D. Eva, C. Evans, M. Foster, S. Fujita, D. Gallagher, 
A. Gaspers, P. Gentry, S. Giddens, J. Graw, M. Hanna, A. Haralson, M. Henderson, D. Herhager, J. Hill, 
S. Horacek, M. Huisjen, S. Hunter, J. Jacob, R. Karre, L. Larsen, P. Lien, R. Manning, J. Marriott, D. McConnell, 
R. McIntosh, M. McIntosh, G. Mead, B. Michelson, B. Miller, J. Moorehead, M. Morris, J. Murphy, M. Nelson, 
J. Pacha, I. Patrick, A. Pearl, B. Pett s. Randall,C. Rowland, R. Sandoval, D. Sealman, K. Shelley, J. Simbai, 
L. Smeins,  C. Stewart, G. Taudien, D. Tennant, J. Troeltzsch, B. Unruh, C. Varner, J. Winghart, J. Workman}
 {\footnotesize \bf Rockwell (Detector Arrays)} {\footnotesize \it
 B. Beardwood, J. Huffman, D. Reynolds, D. Seib, M. Stapelbroek, S. Stetson} {\footnotesize \bf
 OCLI (Filters)} {\footnotesize \it S. Corda, B. Dungan, D. Favot, S. Highland, M. Inong, V. Jauregui, C. 
 Kennemore, B. Langley, S. Mansour, R. Mapes, M. Mazzuchi, C. Piazzo.}
 
{\bf Jet Propulsion Laboratory, California Institute of Technology - Project and Science Management, 
Mission Operations:} {\footnotesize \it  D. Achhnani, A. Agrawal, T. Alfery, K. Anderson, J. Arnett, B. Arroyo, 
D. Avila, W. Barboza, M. Bareh, S. Barry, D. Bayard, C. Beichman, M. Beltran, R. Bennett, P. Beyer, 
K. Bilby, D. Bliss, G. Bonfiglio, M. Bothwell, J. Bottenfield, D. Boussalis, C. Boyles, M. Brown, P. Brugarolas, 
R. Bunker, C. Cagle, C. Carrion, J. Casani, E. Cherniack , E. Clark, D. Cole, J. Craft, J. Cruz, 
M. Deutsch, J. Dooley, S. Dekany,
R. Dumas,  M. Ebersole, C. Elachi, W. Ellery, D. Elliott, K. Erickson, J. Evans, J. Fanson, T. Feehan, R. Fragoso, 
L. Francis,  M. Gallagher, G. Ganapathi,  M. Garcia, T. Gavin, S. Giacoma, J. Gilbert, L. Gilliam, C. Glazer, P. Gluck, 
V. Gorjian, G. Greanias, C. Guernsey, A. Guerrero, M. Hashemi, G. Havens, C. Hidalgo, E. Higa, J. Hodder, 
H. Hotz, W. Hu, J. Hunt-Jr., D. Hurley, J. Ibanez, K. Jin, M. Johansen, M. Jones, J. Kahr, B. Kang, 
P. Kaskiewicz, D. Kern, T. Kia, M. Kline, B. Korechoff, P. Kwan, J. Kwok, H. Kwong-Fu, M. Larson,, M. Leeds, 
R. Lineaweaver, S. Linick, P. Lock, W. Lombard, S. Long, T. Luchik, J. Lumsden, M. Lysek, G. Macala, 
S. Macenka, N. Mainland, E. Martinez, M. Mcauley, J. Mehta, P. Menon, R. Miller, C. Miyamoto,  
W. Moore, F. Morales, R. Morris, A. Nakata, B. Naron, A. Nash, D. Nichols, 
M. Osmolovsky, K. Owen-Mankovich, K. Patel, 
S. Peer, N. Portugues, D. Potts, S. Ramsey, S. Rangel, R. Reid, J. Reimer, E. Rice, D. Rockey, E. Romana, 
C. Rondeau, A. Sanders, M. Sarrel, V. Scarffe, T. Scharton, C. Scott, P. K. Sharma, T. Shaw, D. Shebel,  
J. Short, C. Simon, B. Smith, R. Smith, P. Sorci, T. Specht, R. Spehalski, G. Squibb, S. Stanboli, K. Stapelfeldt, 
D. Stern, K. Stowers, J. Stultz, M. Tankenson, N. Thomas, R. Thomas, F. Tolivar, R. Torres, R. Tung, N. Vandermey, 
P. Varghese, M. Vogt, V. Voskanian, B. Waggoner, L. Wainio, T. Weise, J. Weiss, K. Weld, 
R. Wilson, M. Winters, S. Wissler, G. Yankura, K. Yetter}

{\bf Lockheed-Martin - Spacecraft, Systems Engineering, Spacecraft Operations:} {\footnotesize \it B. Adams, 
J. Akbarzadeh, K. Aline, T. Alt, G. Andersen, J. Arends, F. Arioli, A. Auyeung, D. Bell, R. Bell, F. Bennett, 
J. Bennett, M. Berning, H. Betts,M. Billian, 
S. Broadhead, B. Bocz, G. Bollendonk, N. Bossio, P. Boyle, T. Bridges, 
C. Brink, R. Brookner, J. Brunton, D. Bucher, M. Burrack, R. Caffrey, S. Carmer, P. Carney, T. Carpenter, 
R. Castro, J. Cattrysse,J. Cernac, G. Cesarone, K. Chan, C. Chang, M. Chuang, D. Chenette, A. Chopra, 
Z. Chou, W. Christensen, K. Chu, W. Clark, J. Clayton, S. Cleland, W. Clements, C. Colborn, 
A. Cooprider, B. Corwin, B. Costanzo, D. Cortes, M. Cox, M. Cox, J. Coyne, S. Curtin, 
G. Dankiewicz, C. Darr, J. Dates, J. Day, S. DeBrock, T. Decker, R. Defoe, J. Delavan, G. Delezynski, 
J. Delk, B. Dempsey, R. Dodder, T. Dougherty, H. Drosdat, G. Du, B. Dudginski, M. Dunn, 
R. Dunn, M. Dunnam, D. Durant, D. Eckart, B. Edwards, M. Effertz, L. Ellis, P. Emig, N. Etling, 
M. Etz, N. Fernando, C. Figge, R. Finch, S. Finnell, A. Fisher, M. Fisher, P. Fleming, D. Ford, 
K. Foster, J. Frakes, P. Frankel, D. Fulton, P. Galli, D. Garcia, M. Gardner, B. Garner, S. Gaskin, 
S. Gasner, M. Geil, E. Georgieva, T. Gibson, B. Goddard, M. Gonzalez, D. Goold, D. Graves, 
S. Gray, I. Grimm, J. Grinwis, M. Gronet, R. Grubic, S. Guyer, M. Haggard, J. Harrison, G. Hauser, 
C. Hayashi, P. Headley, W. Hegarty, S. Heires, J. Herrerias, D. Hirsch, K. Hooper, J. Horwath, 
S. Housten, D.Howell, L. Huff, G. Idemoto, B. Jackson, K. Janeiro, K. Johnson, M. Johnson, 
R. Kaiser, P. Kallemeyn, G. Kang, R. Kasuda, M. Kawasaki, B. Keeney, J. Kenworthy, C. King, 
A. Klavins, K. Klein, C. Klien, P. Klier, C. Koch, L. Koch, D. Koide, R. Kriegbaum, J. Kuchera, 
J. Ladewig, D. Lance, M. Lang, K. Lauffer, A. Lee, E. Lee, J. Lee, R. Lee, D. Leister, K. Loar, 
A. Lott, C. Love, N. Iyengar, P. Ma, A. Magallanes, A. Mainzer, T. Maloney, S. Mar, B. Marquardt, 
M. Martin, G. Mason, R. Maxwell, R. May, G. McAllister, S. McElheny, M. McGee, J. McGowan, 
D. McKinney, A. McMechen, E. Merlo, C. Mifsud, J. Miles, S. Miller, A. Minter, C. Miran, S. Mittal, 
R. Mock, R. Mock, J. Montgomery, J. Moore, H. Mora, M. Moradia, L. Morales, R. Morales, 
G. Morison, J. Mota, F. Moules, S. Mumaw, L. Naes, A. Nalbandian, J. Nelson, L. Nenoff, 
J. Neuman, D. Nguyen, K. Nguyen, T. Nguyen, D. Niebur, D. Nishimura, M. Ochs, T. Oliver, 
J. Oo, J. Ortiz, G. Pace, L. Padgett, N. Page, G. Painter, H. Pandya, L. Pappas, N. Pemberton, 
R. Peterson, H. Phan, L. Phan, J. Pine, R. Poling, R. Potash, D. Radtke, W. Ramos, T. Ransom, 
M. Ratajczyk, D. Read, S. Ready, M. Rich, R. Richey, H. Rizvi, C. Rollin, C. Rudy, M. Rugebregt, 
R. Russek, B. Sable, C. Sandwick, M. Santos, N. Schieler, J. Schirle, G. Schlueter, M. Schmitzer, 
E. Sedivy, R. Seeders, S. Selover, R. Shaw, F. Sheetz, D. Shelton, R. Sherman, T. Sherrill, 
O. Short, R. Sison, B. Smith, F. Smith, S. Smith, B. Sotak, S. Spath, J. St. Pierre, K. Starnes, 
K. Stowers, J. Straetker, T. Stretch, S. Sulak, W. Sun, D. Swanson, C. Tatro, M. Tebo, D. Telford, 
A. Tessaro, J. Tietz, D. Tenerelli, J. Tolomeo, S. Toro-Allen, J. Tousseau, R. Traber, M. Tran, 
P. Travis, K. Uselman, S. Utke, N. Vadlamudi, R. VanBezooijen, J. Vantuno, R. Vasquez, 
G. Vergho, C. Voth, B. Vu, P. Wagner, M. White, M. Whitten, J. Wood, C. Worthley, D. Wright, C. Yanari, 
L. Yeaman, D. Zempel, S. Zeppa}

{\bf NASA-Ames Research Center - Project Management Through 1989:} {\footnotesize \it W. Brooks, 
P. Davis, A. Dinger, L. Manning, R. Melugin, J. Murphy, R. Ramos, C. Wiltsee, F. Witteborn, L. Young
% ...
}

\begin{figure} 
\includegraphics[height=6.0in]{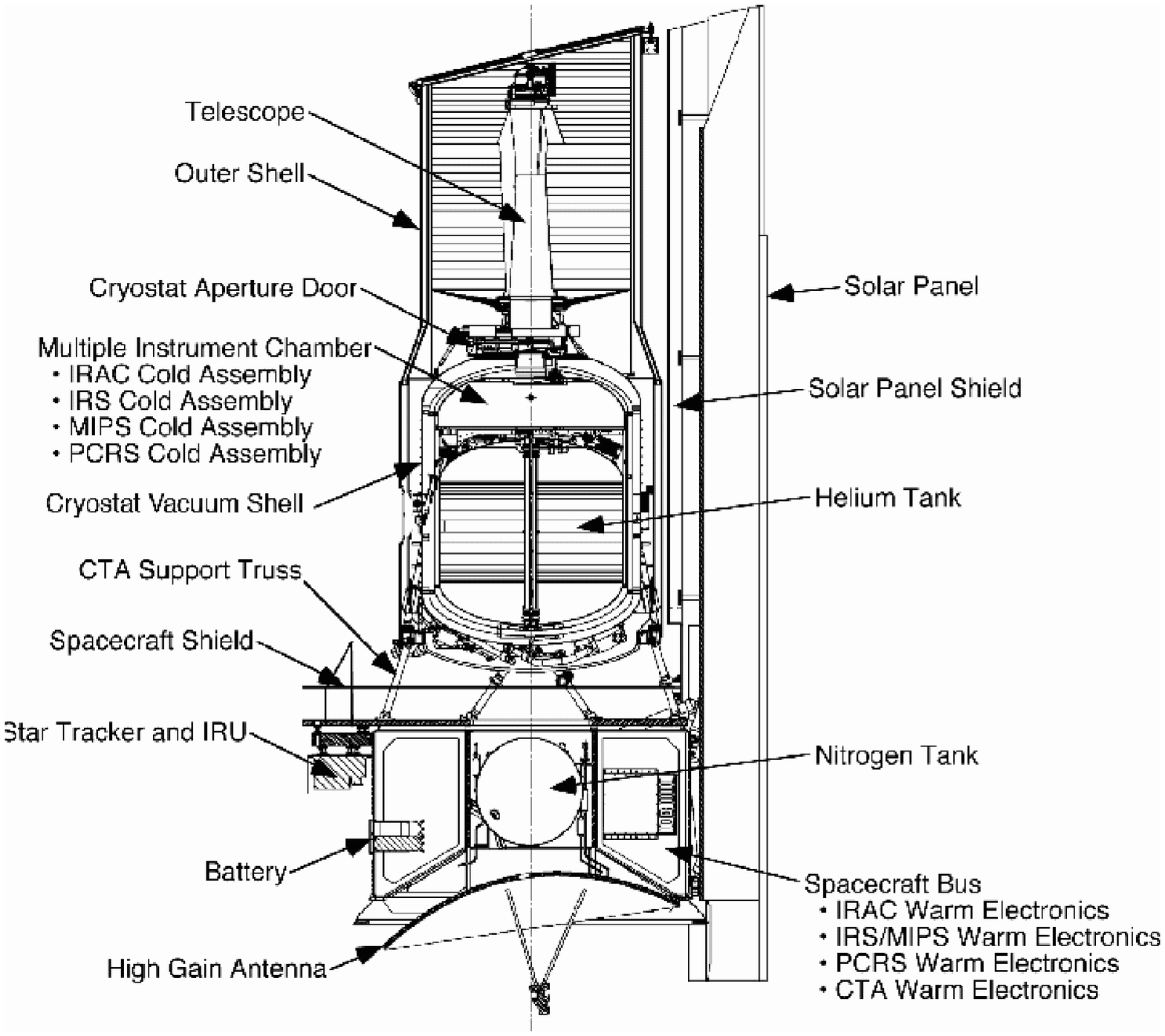}

\caption{Cutaway view of the Spitzer Space Telescope
flight hardware.  The observatory is approximately 4.5m high and 2.1m in diameter.
In this figure the dust cover is shown prior to its ejection approximately five
days after launch. \label{figure1}}
\end{figure}

\begin{figure} 
\includegraphics[height=6.0in]{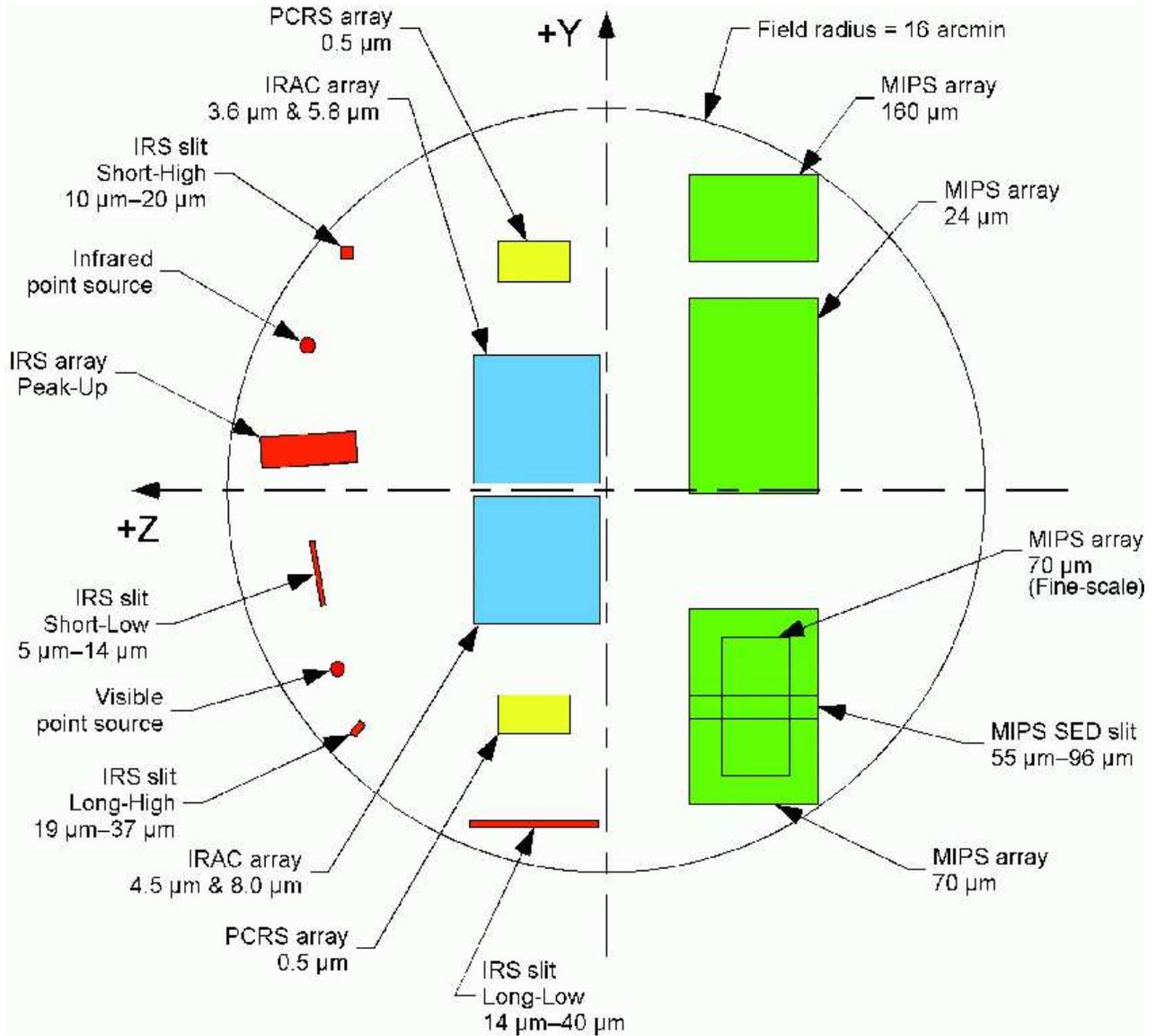}

\caption{The SST focal plane layout as seen looking down
from the telescope aperture. In this coordinate system the $+Z$ 
direction points toward the
sun. In addition to the science instrument apertures, the figure 
shows the location of the
two Pointing Calibration Reference Sensor (PCRS) arrays and the two 
point sources that were
used for ground-based focus checks and focal-plane mapping.
The MIPS apertures appear rectangular because the scan mirror accesses an area
larger than the instantaneous footprint of the aray; the position of the SED slit and the fine
scale array are shown schematically.\label{figure2}}
\end{figure}

\begin{figure} 
\includegraphics[height=5.0in]{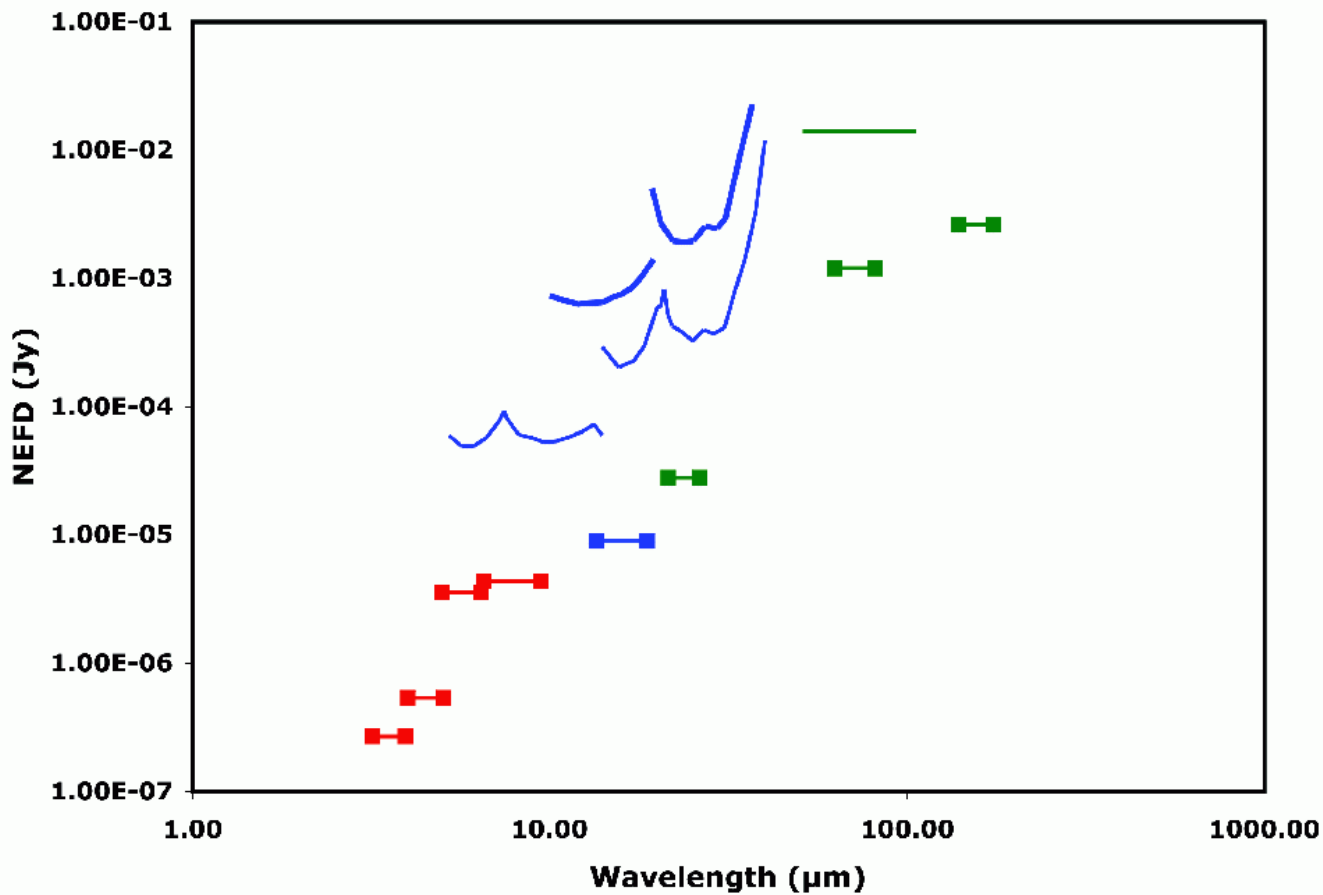}

\caption{The point-source sensitivity and wavelength 
coverage of the Spitzer
science instruments. All data are for 500~second integration times 
and are 1-$\sigma$. In most
cases the realized sensitivity can depend strongly on the zodiacal 
emission level,
background confusion, and for the MIPS 70\,$\mu$m array, which half 
of the array is being
used. The data presented here are for the most optimal locations on 
the sky and the most
sensitive half of the MIPS 70\,$\mu$m array. The red lines  with filled-square
endpoints show the IRAC sensitivity and wavelength coverage. The blue lines
denote the IRS, with the blue line  with the filled square end points 
indicating the IRS blue peakup array when used for imaging. The IRS red 
peakup array performance is not indicated on this
plot as it is very similar to the MIPS 24\,$\mu$m channel in its 
wavelength coverage but is not as sensitive. The blue lines without endpoints indicate
the IRS spectral coverage and sensitivity, with the thicker lines denoting the high-resolution
modules and the thinner lines the low-resolution module. The green lines with the filled
end points indicate the MIPS sensitivity and wavelength coverage. The green line without 
endpoints indicates the predicted MIPS SED performance, although since this 
operational mode has not been commissioned yet there is only incomplete flight data available on its sensitivity. \label{figure3}}
\end{figure}

\clearpage

\begin{deluxetable}{lc}
\tablewidth{0pt}
\tablecaption{Top-Level Observatory Parameters}
\tablehead{
\colhead{Parameter}&\colhead{Value}}
\startdata
Total Observatory mass at launch 	&861 kg\\
Dimensions (height x diameter)&	4.5\,m  x 2.1\,m\\
Average operating power	&375\,W\\
Solar array generating capacity at launch	&500\,W\\
Nitrogen reaction control gas at launch	&15.59\,kg\\
Estimated reaction control gas lifetime	&17 years\\
Mass memory capacity	&2\,GB\\
Telescope primary diameter	&0.85\,m\\
Telescope central obscuration	&14.2\%\\
Superfluid helium at launch	&337\,l\\
Estimated nominal cryogenic lifetime	&5.3 years\\
Data transmission rate (high-gain antenna up to 0.58\,AU from Earth) 
	&2.2\,Mb\,s$^{-1}$\\
Command communication rate &	2\,kbps\\
\enddata
\end{deluxetable}

\clearpage

\begin{deluxetable}{lccc}
\tablewidth{0pt}
\tablecaption{Infrared Array Camera (IRAC)---Four Channel Infrared Imager
\protect\\ PI: G.~G.~Fazio, SAO; Instrument built at NASA-Goddard }
\tablehead{
\colhead{Channel}&\colhead{Wavelength [$\mu$m]}&\colhead{Field of View [arc-min]}&
\colhead{Detector}}
\startdata
1	&3.19--3.94	&5.2 x 5.2	&256 x 256 InSb\\
2	&4.00--5.02	&5.2 x 5.2	&256 x 256 InSb\\
3	&4.98--6.41	&5.2 x 5.2	&256 x 256 Si:As\\
4	&6.45--9.34	&5.2 x 5.2	&256 x 256 Si:As\\
\enddata
\end{deluxetable}

\clearpage

\begin{deluxetable}{lccc}
\tabletypesize{\small}
\tablewidth{0pt}
\tablecaption{Infrared Spectrograph (IRS)---Low/Moderate Resolution 
Spectrometer \protect\\PI: J.~R.~Houck, Cornell University; Instrument built at Ball Aerospace}
\tablehead{
\colhead{Module} & \colhead{Wavelength Range} &
\colhead{Slit Dimensions} & \colhead{Spectral Resolution } 
\\ \colhead{} & \colhead{[$\mu$m]} & \colhead{ [arc-sec]} 
& \colhead{$\lambda/\Delta\lambda$}}
\startdata
Short-Low	&5.2--7.7 second-order& 3.6 x 57 & 80--128\\
\nodata&7.4--14.5 first-order	& 3.7 x 57 & 64--128\\
Long-Low	&14.0--21.3 second-order &10.5 x 168 &80--128\\
\nodata&19.5--38.0 first-order	&10.7 x 168 &64--128\\
Short-High	&9.9--19.6	& 4.7 x 11.3	& $\sim$600\\
Long-High&18.7--37.2	&11.1 x 22.3	&$\sim$600\\
Peakup Array (Blue)	&13.3--18.7	&56 x 80&	3\\
Peakup Array (Red) &18.5--26.0 &54 x 82 &3\\
\enddata
\end{deluxetable}

\clearpage

\begin{deluxetable}{lccc}
\tabletypesize{\small}
\tablewidth{0pt}
\tablecaption{Multiband Infrared Photometer for Spitzer (MIPS)---Far-Infrared 
Imager and Spectral Energy Distribution (SED) Photometer 
\protect\\PI: G.~H.~Rieke, Arizona; Instrument built at Ball Aerospace}
\tablehead{
\colhead{Band Identification} & \colhead{Wavelength Range} &
\colhead{Field of View [arc-min] }&	\colhead{Detector Array}
\\ \colhead{} & \colhead{ [$\mu$m]} & \colhead{ [arc-min]} & \colhead{}}
\startdata
24~$\mu$m	&21.5---26.2	& 5.4 x 5.4	&128 x 128 Si:As\\
70~$\mu$m	&62.5---81.5	& 5.2 x 5.2\tablenotemark{a}	&32 x 
32 Ge:Ga\\
160~$\mu$m	&139.5---174.5\tablenotemark{b}	&5.3 x 0.5	&2 x 
20 Stressed Ge:Ga\\
SED	 ($\lambda/\Delta\lambda = 15$--25)&51--106	&2.7 x 0.34 
	&32 x 24 Ge:Ga\\
\enddata

\tablenotetext{a}{The MIPS 70 \micron\ array consists of two 5.2\,arc-min x 
2.6\,arc-min halves. Due to a
problem in the cold cabling one of the halves has significantly worse 
sensitivity than the
other.}

\tablenotetext{b}{The MIPS 160 \micron\ channel has a short-wavelength 
filter leak that admits some
1.6 \micron\ light that must be accounted for when observing blue objects. 
See the Spitzer Science Center web-site
for more information.}

\end{deluxetable}


\clearpage

\begin{thebibliography}{}

\bibitem[Bell Burnell, Davies, \& Stobie (1992)]{bel92} Bell Burnell, S. J., Davies, J. K., \&
Stobie, R. S. 1992, Next Generation Infrared Space Observatory, Dordrecht, Kluwer.

\bibitem[Kwok (1993)]{kwo93}Kwok, J. H. 1993, in Advances in Astronautical
Sciences, ed. V. J. Modi et al. (American Astronautical Society), Volume 35, paper 93-651. 

\bibitem[Fazio et al.(2004)]{faz04} Fazio, G.,  et al. 2004, \apjs,
 this volume

\bibitem[Finley, Hopkins, \& Schweickart (2004)]{fin04} Finley, P. T.,Hopkins, R. A., 
\& Schweickart, R.B.
2004, \procspie, 5487-02


\bibitem[Gehrz et al.(2004)]{geh04} Gehrz, R. D., et al. 2004, \procspie, 5487-86

\bibitem[Hoffmann et al.(2004)]{hof04} Hoffmann, W. F.,  Hora, J. 
L., Mentzell, J. E.,
Marx, C. T., Eisenhardt, P. R., Carey, S. J., \& Megeath, S. T. 
2004, \procspie, 5487-88

\bibitem[Houck et al.(2004)]{hou04} Houck, J. R., et al. 2004, \apjs.
 this volume

\bibitem[Lysek et al.(1995)]{lys95} Lysek, M. J., Israelsson, U. E., Garcia, R. D., \&
Luchik,  T. S 1995, in Advances in Cryogenic Engineering, ed. P. Kittel, 
(Plenum Publishing) Volume 41, 1143

\bibitem[Rieke et al.(2004)]{rie04} Rieke, G., et al. 2004, \apjs,
 this volume

\end{thebibliography}
\end{document}